\documentclass[draft]{aipproc}
\layoutstyle{6x9} 

\def\lqcd{\Lambda_{\rm QCD}}
\def\jp{J/\psi}
\def\psip#1{\psi_{\mathbf{#1}}}
\def\chip#1{\chi_{\mathbf{#1}}}
\def\bsigma{\mathbf \sigma}

\begin{document}

\title[{\sl 9th International Symposium on Heavy Flavor Physics}]
      {Production \& Decay of Quarkonium}


\author{Sean Fleming}{
  address={Dept. of Physics Carnegie Mellon University, Pittsburgh PA
  15213, USA},
  email={fleming@kayenta.phys.cmu.edu},
  thanks={}
}

\copyrightyear  {2001}


\begin{abstract}
In this talk I review NRQCD predictions for the production of
charmonium at the Tevatron. After a quick presentation of the NRQCD
factorization formalism for production and decay I review some old
results and discuss how they compare to recent data. Following this I
discuss some recent work done with Adam Leibovich and Ira Rothstein.
\end{abstract}

\date{\today}


\maketitle

\section{Introduction}

Heavy quarkonia have proven fruitful in helping us gain a better
understanding of QCD. Early theoretical analyses of quarkonia decay
were based on the color-singlet model (CSM)~\cite{Buchmuller:zf}. The
underlying assumption of this model is that the heavy-quark--antiquark
pair has the same quantum numbers as the quarkonium meson. (For
example the $b\bar{b}$ that forms an $\Upsilon$ must be in a
color-singlet ${}^3S_1$ configuration.) One consequence of such a
restrictive assumption is that theoretical predictions based on the
CSM are simple, depending on only one nonperturbative parameter.
However, the CSM does not provide a systematic approach to studying
quarkonium. This is clear in $P$-wave decays where infrared
divergences signal the breakdown of the CSM~\cite{Bodwin:1992ye}.

In order to systematically study nonrelativistic systems long distance
physics needs to be separated from short distance physics. This can be
accomplished with a proper effective field theory which provides a
power counting that determines relevant operators. In most effective
theories the power counting is based upon dimensional analysis,
however, for non-relativistic QCD (NRQCD)~\cite{bbl,lmr} this is not
the case. Instead, it is an expansion in the parameter $v$, the
relative velocity of the heavy quarks. This leads to the result that
operators of the same dimension may be of different orders in the
power counting.  Inclusive decay rates and production cross sections
are now understood within the framework of NRQCD
factorization~\cite{bbl}, where decay rates and production cross
sections are predicted in a systematic double expansion in $\alpha_s$ and
$v$. These predictions have met with varying degrees of success.

In the first half of my talk I give a quick review of NRQCD, and the
NRQCD factorization formalism for production and decay. This gives the
background needed to understand theoretical predictions for production
and decay of $J/\psi$ and $\psi'$. I do not attempt to review all of
the predictions, instead I focus on the transverse momentum
distribution of $J/\psi$ and $\psi'$ at the Tevatron. For the reason 
that here lies the earliest success and the greatest challenge to
NRQCD factorization. In Ref.~\cite{Braaten:1994vv} the $\psi'$
`anomaly' (a factor of 30 discrepancy between the CSM prediction and
date) was resolved using NRQCD. However, the initial data on the
polarization of these states at large transverse
momentum~\cite{Affolder:2000nn} seems to be at odds with the NRQCD
prediction.

In the second half of the talk I will discuss a possible resolution to
the charmonium polarization puzzle at the Tevatron. This is based on
recent work~\cite{Fleming:2000ib} with Ira Rothstein and Adam
Leibovich wherein we propose an alternative power counting for
charmonium. I do not present all the pieces of evidence which seems to
tell us that the effective field theory which best describes the
$J/\psi$ system may not be the same theory which best describes the
$\Upsilon$. For that the reader is directed to the literature. I
merely give a quick review of the new power counting, and then proceed
to discuss how this changes the predictions for $J/\psi$ and $\psi'$
polarization at the Tevatron.

\section{NRQCD}

The power counting depends upon the relative size of the four scales
$(m,mv,mv^2,\lqcd)$. If we take $m>mv> mv^2\simeq \lqcd$ the
bound state dynamics will be dominated by exchange of Coulombic gluons
with $(E\simeq mv^2,\vec{p}=m\vec{v})$.  This hierarchy has been
assumed in the NRQCD calculation of production and decay rates and is
probably a reasonable choice for the $\Upsilon$ system, where
$mv \sim 1.5 {\rm\ GeV}$. However, whether or not it is correct for
the $\jp$, where $mv \sim 700 {\rm\ MeV}$ remains to be seen.

The power counting can be established in a myriad of different ways.
Here I will follow the construction of \cite{lmr}, which I now briefly
review. There are three relevant gluonic modes \cite{Beneke:1997zp}:
the Coulombic $(mv^2,mv)$, soft ($mv,mv$) and ultrasoft
$(mv^2,mv^2)$. The soft and Coulombic modes can be integrated out
leaving only ultrasoft propagating gluons. In the process of
integrating out these modes large momenta must be removed from the
quark field. This is accomplished by rescaling the heavy quark fields
by a factor of $\exp(i {\bf p}\cdot {\bf x})$ and labeling them by
their three momentum ${\bf p}$.  The ultrasoft gluon can only change
residual momenta and not labels on fields.  This is analogous to HQET,
where the four-velocity labels the fields and the nonperturbative
gluons only change the residual momenta \cite{Manohar:dt}.  This
rescaling must also be done for soft gluon fields
\cite{Griesshammer:1997wz} which, while they cannot show up in
external states, do show up in the Lagrangian. After this rescaling a
matching calculation leads to the following tree level Lagrangian
\cite{lmr}
\begin{eqnarray}\label{nrqcd:1}
{\cal L} &=&  
  \sum_{\mathbf p}
 \psip p ^\dagger 
  \Biggl\{ i D^0 - {{\mathbf p}^2 \over 2 m} \Biggr\} 
 \psip p   
 - 4 \pi \alpha_s \sum_{q,q^\prime{\mathbf p},{\mathbf p^\prime}}
 \Bigg\{{1\over q^0} \psip {p^\prime} ^\dagger  
  \left[A^0_{q^\prime},A^0_q \right]
 \psip p 
\nonumber \\
&&  +{g^{\nu 0} (q^\prime-p+p^\prime)^\mu -
        g^{\mu 0} (q-p+p^\prime)^\nu + 
        g^{\mu\nu}(q-q^\prime)^0 \over 
        ( {\bf p}^\prime-{\bf p} )^2}
    \psip {p^\prime} ^\dagger  \left[A^\nu_{q^\prime},A^\mu_q \right]
    \psip p \Bigg\}
\nonumber \\
&&\qquad\qquad + 
\psi \leftrightarrow \chi,\ T \leftrightarrow \bar T + 
\sum_{{\bf p},{\bf q}}
  {4 \pi \alpha_s  \over  ( {\bf p}-{\bf q} )^2} 
  \psip q ^\dagger T^A \psip p \chip {-q}^\dagger  \bar T^A \chip {-p}
  + \ldots
\end{eqnarray}
where we have retained the lowest order terms in each sector of the
theory. The matrices $T^A$ and $\bar T^A$ are the color matrices for
the $\bf 3$ and $\bf \bar 3$ representations, respectively. Note the
last term is the Coulomb potential, which is leading order and must be
resummed in the four-quark sector, while the other non-local
interactions arise from soft gluon scattering.

All the operators in the Lagrangian have a definite scaling in $v$,
and the spin symmetry, which will play such a crucial role in the
polarization predictions, is manifest. The two subleading interactions
which will dominate my discussion are the electric dipole ($E1$)
\begin{equation}
{{\cal L}_{E1}}=\psip p  ^\dagger 
\frac{{\mathbf p}}{m}\cdot {\mathbf A} \psip p \,,
\end{equation}  
and the magnetic dipole ($M1$)
\begin{equation}
{{\cal L}_{M1}}= c_F\, g \psip p ^\dagger {{\bf \bsigma \cdot B} \over 2
m}\psip p \,.
\end{equation} 
The $E1$ interaction is down by a factor of $v$ while the $M1$ is down
by a factor of $v^2$. 

\section{NRQCD Factorization Formalism}

In the NRQCD factorization formalism developed in Ref.~\cite{bbl}
decay rates and production cross sections are written as a sum of
products of Wilson coefficients encoding short distance physics and
NRQCD matrix elements describing long distance physics. In this
formalism a general decay process is written as
\begin{equation}
 \Gamma_{J/\psi}=\sum  C_{^{2S+1} L_J}(m,\alpha_s)\langle \psi \mid
  O^{(1,8)}(^{2S+1}L_J) \mid \psi \rangle.
\end{equation}
The matrix element represents the long distance part of the rate and
may be thought of as the probability of finding the heavy quarks in
the relative state $n$, while the coefficient
$C_{^{2S+1}L_J}(m,\alpha_s)$ is a short distance quantity calculable
in perturbation theory. The sum over operators may be truncated as an
expansion in the relative velocity $v$.  Similarly, production cross
sections may be written as
\begin{equation} 
\label{prodrate} 
d\sigma =\sum_n d\sigma_{i+j\rightarrow Q\bar{Q}[n]+X}\langle 0 \mid
O^H_n 
\mid 0 \rangle. 
\end{equation} 
Here $d\sigma_{i+j\rightarrow Q\bar{Q}[n]+X}$ is the short distance
cross section for a reaction involving two partons, $i$ and $j$, in
the initial state, and two heavy quarks in a final state, labeled by
$n$, plus $X$.  This part of the process is calculable in perturbation
theory, up to possible structure functions in the initial state. The
production matrix elements, which differ from those used in the decay
processes, describe the probability of the short distance pair in the
state $n$ to hadronize, inclusively, into the state of interest.  The
relative size of the matrix elements in the sum are again fixed by the
power counting which we will discuss in more detail below.

The formalism for decays is on the same footing as the operator
product expansion (OPE) for non-leptonic decays of heavy quarks, while
the production formalism assumes factorization, which is only proven,
and in some applications of production this is not even the case, in
perturbation theory~\footnote{For a discussion of factorization in
NRQCD see Refs.~\cite{bbl,Beneke:1996tk}.}~\cite{Collins:gx}. The
trustworthiness of factorization depends upon the particular
application. I have reviewed these results here to emphasize the point
that when the theory is tested one is really testing both the
factorization hypothesis as well the validity of the effective theory
as applied to the $J/\psi$ system.  Thus, care must be taken in
assigning blame when theoretical predictions do not agree with data.

\section{Charmonium production at the Tevatron}

Having introduced NRQCD and the factorization formalism I will now
turn to theoretical predictions of $J/\psi$ and $\psi'$ production at
the Tevatron. The leading order in $v$ contribution to $J/\psi$
production is through the color-singlet matrix element $\langle
O_1^\psi(^3S_1)\rangle$, since the quantum numbers of the short
distance quark pair matches those of the final state.  All other
matrix elements need insertions of operators into time ordered
products to give a non-zero result, and are therefore suppressed
compared to the color-singlet matrix element above.  For instance, the
matrix element $\langle O_8^\psi(^1S_0)\rangle$ vanishes at leading
order.  The first non-vanishing contribution comes from the insertion
of two $M1$ operators into time ordered products, thus giving a $v^4$
suppression.  The scaling of the relevant matrix elements for $\psi$
production are shown in Table~\ref{psiMEscaling} under NRQCD$_b$ (for
reasons which I will explain later). It appears from just the $v$
counting that only the color-singlet contribution is
important. However, other contributions can be enhanced by
kinematic factors. At large transverse momentum, fragmentation type
production dominates~\cite{Braaten:1993rw}, and only the $\langle
O_8^\psi(^3S_1)\rangle$ contribution is important.  Without the
color-octet contributions ({\it i.e.}, the Color-Singlet Model), the
theory is below experiment by about a factor of 30.  By adding the
color-octet contribution the fit to the data is very good
\cite{Braaten:1994vv}. 
\begin {table}[t]
\begin {tabular}{ccccc}
\\
\hline
\\
 & $\langle O_1^\psi(^3S_1)\rangle$  
 & $\langle O_8^\psi(^3S_1)\rangle$ 
 & $\langle O_8^\psi(^1S_0)\rangle$ 
 & $\langle O_8^\psi(^3P_0)\rangle$ \\\\
\hline
\\
NRQCD$_b$  & $v^0$ &  $v^4$ &  $v^4$ &  $v^4$ \\\\
NRQCD$_c$  & $(\lqcd/m_c)^0$ 
        & $(\lqcd/m_c)^4$ 
        & $(\lqcd/m_c)^2$ 
        & $(\lqcd/m_c)^4$ \\\\
\hline
\end {tabular}
\caption{Scaling of matrix elements relevant for $\psi$ production in
NRQCD$_b$ and NRQCD$_c$.}
\label{psiMEscaling}
\end {table}

Once the color-octet matrix elements are fit to the unpolarized date
it is possible to make a parameter free prediction for the
polarization of $J/\psi$ and $\psi'$ at the Tevatron: they are
predicted to be transversely polarized at large $p_T$. This is
because at large transverse momentum, the dominant production
mechanism is through fragmentation from a nearly on shell gluon to the
octet $^3S_1$ state.  The quark pair inherits the polarization of the
fragmenting gluon, and is thus transversely polarized~\cite{Cho:1994ih}. The
leading order transition to the final state goes via two $E1$, spin
preserving, gluon emissions.  Higher order perturbative fragmentation
contributions~\cite{Beneke:1995yb}, fusion diagrams
\cite{Beneke:1996yw,Leibovich:1996pa}, and feed-down for the
$J/\psi$~\cite{Braaten:1999qk} dilute the polarization some, but the
prediction still holds that as $p_T$ increases so should the
transverse polarization.  Indeed, for the $\psi^\prime$, at large
$p_T\gg m_c$, we expect nearly pure transverse polarization.  This
prediction seems to be at odds with the initial data which seems to
suggest that the $J/\psi$ and the $\psi'$ are unpolarized or slightly
longitudinally polarized as $p_T$
increases~\cite{Affolder:2000nn}.\footnote{The data still has rather
large error bars, so we should withhold judgment until the statistics
improves.} If, after the statistics improve, this trend continues one
is left with two obvious possibilities: 1) The power counting of NRQCD
does not apply to the $J/\psi$ system. 2) Factorization is violated
``badly'', meaning that there are large power corrections.

\section{NRQCD$_c$}

In this section I discuss the work done with Adam Leibovich and Ira
Rothstein in Ref.\cite{Fleming:2000ib}. In that work we marshaled
evidence that the NRQCD power counting might not apply to the $J/\psi$
system. We did not consider the second possibility mentioned at the end
of the previous section: that factorization is violated. I will not be
considering this possibility either.

The standard NRQCD methodology, which is based upon the hierarchy
$m>mv> mv^2\simeq \lqcd$, has been applied to the $J/\psi$ as well as
the $\Upsilon$ systems. While it seems quite reasonable to apply this
power counting to the $\Upsilon$ system, it is not clear that it
should apply to the $J/\psi$ system. Indeed, I believe that the data
is hinting toward the possibility that a new power counting is called
for in the charmed system.

If NRQCD does not apply to the $J/\psi$ system, then one must ask: is
there another effective theory which does correctly describe the
$J/\psi$?  One good reason to believe that such a theory does exist is
that NRQCD, as formulated, does correctly predict the ratios of decay
amplitudes for exclusive radiative decays. Using spin symmetry the
authors of \cite{Cho:1994ih} made the following predictions:
\begin{eqnarray}
&&\Gamma(\chi_{c0}\rightarrow J/\psi+\gamma):
\Gamma(\chi_{c1}\rightarrow J/\psi+\gamma):\Gamma(\chi_{c2}\rightarrow
J/\psi+\gamma)
:\Gamma(h_{c}\rightarrow \eta_c+\gamma) \nonumber \\
&& =0.095:0.20:0.27:0.44\quad {\rm (theory)}\nonumber \\
&& =0.092\pm 0.041 :0.24\pm 0.04:0.27\pm
   0.03:{\rm unmeasured} \quad {\rm (experiment)}.
\end{eqnarray}
Thus, an alternative formulation of NRQCD must preserve these
predictions yet yield different predictions in other relevant
processes. 

Let me now consider the alternate hierarchy $m>mv\sim \lqcd$.  One
might be tempted to believe that in this case the power counting
should be along the lines of HQET, where the typical energy and
momentum exchanged between the heavy quarks is of order $\lqcd$.
However, this leads to an effective theory which does not correctly
reproduce the infra-red physics. With this power counting, the leading
order Lagrangian would simply be
\begin{equation}
{\cal L_{\rm HQET}}=\psi_v ^\dagger D_0 \psi_v,
\end{equation}
where the fields are now labeled by their four velocity.  This is a
just a theory of time-like Wilson lines (static quarks) which does not
produce any bound state dynamics.  Thus I am forced to conclude that
the typical momentum is of order $\lqcd$, whereas the typical energy
is $\lqcd^2/m$, so that $D^2/(2m)$ is still relevant. I will call
this theory NRQCD$_c$, and will refer to the traditional power
counting as NRQCD$_b$ as I assume that it describes the bottom
system.
 
The power counting of this theory is now along the lines of HQET where
the expansion parameter is $\lqcd/m_Q$.  However the residual energy of
the quarks is order $\lqcd^2/m_Q$, while the residual three momentum
is $\lqcd$.  Thus one must be careful in the power counting to
differentiate between time and spatial derivatives acting on the quark
fields.  As far as the phenomenology is concerned, perhaps the most
important distinction between the power counting in NRQCD$_c$ and
NRQCD$_b$ is that the magnetic and electric gluon transitions are now
of the same order in NRQCD$_c$.  This difference in scaling does not
disturb the successes of the standard NRQCD$_b$ formulation but does
seem help in some of its shortcomings.

\section{NRQCD$_c$ predictions}

The relative size of the different matrix elements change in
NRQCD$_c$.  In particular, the $M1$ transition is now the same order
as the $E1$ transition.  The new scaling is shown in
Table~\ref{psiMEscaling}.~\footnote{These results reproduce those
given in \cite{Beneke:1997av} when $\lambda$ is taken to be 1 in this
reference.}  Due to the dominance of fragmentation at large transverse
momentum, we need to include effects up to order $(\lqcd/m_c)^4$,
since the $\langle O_8^\psi(^3S_1)\rangle$ matrix element will still
dominate at large $p_T$.

Is this consistent?  The size of the matrix elements is a clue.
Extraction of the matrix elements uses power counting to limit the
number of channels to include in the fits.  Calculating $J/\psi$ and
$\psi'$ production up to order $(\lqcd/m_c)^4$ in NRQCD$_c$ requires
keeping the same matrix elements as in NRQCD$_b$.  Previous
extractions of the matrix elements only involve the linear combination
\begin{equation}
M_r^\psi = \langle O_8^\psi(^1S_0)\rangle 
   + \frac{r}{m_c^2}\langle O_8^\psi(^3P_J)\rangle, 
\end{equation}
with $r\approx3-3.5$, since the short-distance rates have similar size
and shape.  In the new power-counting, I can just drop the
contribution from $\langle O_8^\psi(^3P_J)\rangle$, since it is down
by $(\lqcd/m_c)^2\sim 1/10$ compared to $\langle
O_8^\psi(^1S_0)\rangle$.  It is the same order as $\langle
O_8^\psi(^3S_1)\rangle$, but is not kinematically enhanced by
fragmentation effects.  The extraction from \cite{Braaten:1999qk}
would then give for the $J/\psi$ and $\psi'$ matrix elements
\begin{eqnarray}
\langle O_8^{\jp}(^1S_0)\rangle:\langle O_8^{\jp}(^3S_1)\rangle
   &=& (6.6 \pm 0.7) \times 10^{-2} :(3.9 \pm 0.7) \times 10^{-3}
   \approx 17 : 1 \,, \nonumber\\
\langle O_8^{\psi'}(^1S_0)\rangle:\langle O_8^{\psi'}(^3S_1)\rangle
   &=& (7.8 \pm 3.6) \times 10^{-3} :(3.7 \pm 0.9) \times 10^{-3}
   \approx 2 : 1.
\end{eqnarray}
Other extractions have various values of the hierarchy, ranging from
$3:1$ to $20:1$ \cite{hier}.  While the relation of the color-octet
matrix elements in the $\jp$ system is indeed in agreement with the
NRQCD$_c$ power counting, the $\psi'$ does not look to be
hierarchical. However, it should be noted that the statistical errors
in the $\psi^\prime$ extraction, quoted above, are quite large.
Furthermore, there are also large uncertainties introduced in the
parton distribution function.  The above ratios used the CTEQ5L parton
distribution functions.  If we take the central values from
\cite{Braaten:1999qk} for the MRST98LO distribution
functions, we find the ratio $3:1$. On the other hand, the $J/\psi$
extraction is much less sensitive to the choice of distribution
function.  Given the statistical and theoretical errors, it clear that
the $\psi^\prime$ ratio is not terribly illuminating.

Let me now consider the  extraction of these color-octet matrix
elements in the $\Upsilon$ sector~\cite{Braaten:2000cm}, where according to
NRQCD$_b$ power counting there is should be no hierarchy:
\begin{eqnarray}
\langle O_8^{\Upsilon(3S)}(^1S_0)\rangle:
\langle O_8^{\Upsilon(3S)}(^3S_1)\rangle &=& 
(5.4 \pm 4.3^{+3.1}_{-2.2}) \times 10^{-2} :
(3.6 \pm 1.9^{+1.8}_{-1.3}) \times 10^{-2}
\nonumber \\ 
&\approx & 1 : 1, \nonumber \\
\langle O_8^{\Upsilon(2S)}(^1S_0)\rangle:
\langle O_8^{\Upsilon(2S)}(^3S_1)\rangle &=& 
(-10.8 \pm 9.7^{-3.4}_{+2.0}) \times 10^{-2} :
(16.4 \pm 5.7^{+7.1}_{-5.1}) \times 10^{-2}
\nonumber \\ 
&\approx & 1 : 1 \nonumber, \\
\langle O_8^{\Upsilon(1S)}(^1S_0)\rangle:
\langle O_8^{\Upsilon(1S)}(^3S_1)\rangle &=& 
(13.6 \pm 6.8^{+10.8}_{-7.5}) \times 10^{-2} :
(2.0 \pm 4.1^{-0.6}_{+0.5}) \times 10^{-2}
\nonumber \\
&\approx & 6 : 1. 
\end{eqnarray}
For the $\Upsilon(3S)$ and $\Upsilon(2S)$ we observe that there is
indeed no hierarchy, while for the $\Upsilon(1S)$ it appears there may
be a hierarchy. However, it is not possible to draw any strong
conclusions from these data because the errors on the extractions are
large. In fact the ratio for the $\Upsilon(1S)$ color-octet matrix
elements is $1:1$ within the one sigma errors. Furthermore, these
matrix elements are those extracted subtracting out the feed down from
the higher states. While phenomenologically it is perfectly reasonable
to define the subtracted matrix elements, I believe that, since the
matrix elements are inclusive, one should not subtract out the feed
down from hadronic decays when checking the power counting. In
principle this subtraction should not change things by orders of
magnitude, but nonetheless it can have a significant effect. Indeed,
if one compares the ratios for inclusive matrix elements, which do not
have the accumulated error, then the ratios come out to be $1:1$, even
for the $\Upsilon(1S)$
\cite{Braaten:2000cm}.

\begin{figure}[ht]
\includegraphics[width=0.47\textwidth]{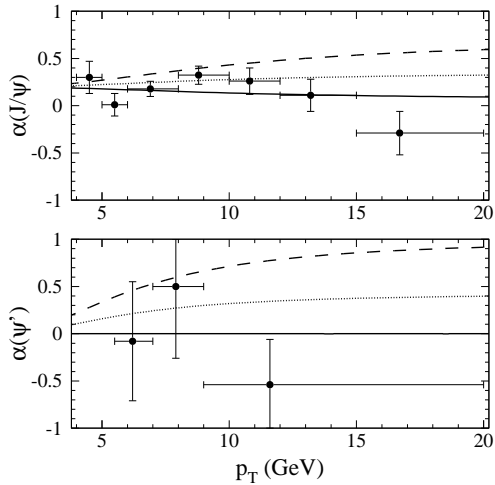}
\caption{Predicted polarization in NRQCD$_c$ for $J/\psi$ and $\psi'$
at the Tevatron as a function of $p_T$.  The three lines correspond to
$R_{M/E}$=(0 (dashed), 1 (dotted), $\infty$ (solid)).  The dashed line
is also the prediction for NRQCD$_b$.}
\label{polar}
\end{figure}

With NRQCD$_c$, the intermediate color-octet $^3S_1$ states hadronize
through the emission of either two $E1$ or $M1$ dipole gluons, at the
same order in $1/m_c$.  Since the magnetic gluons do not preserve
spin, the polarization of $\psi$ produced through the $\langle
O_8^{\psi}(^3S_1)\rangle$ can be greatly diluted.  The net
polarization will depend on the ratio of matrix elements
\begin{eqnarray}
R_{M/E} &=&
\\
&& \hspace{-14pt}\frac{\int \prod_{\ell} d^4x_{\ell}
 \langle 0 \mid T(M_1(x_1)M_1(x_2)\psi^\dagger T^a\sigma_i \chi)
 \,a_H^\dag\,a_H\,T(M_1(x_3)M_1(x_4)\chi^\dagger T^a\sigma_i \psi) 
 \mid 0\rangle}
{\int \prod_{\ell}d^4x_{\ell}  
 \langle 0 \mid T(E_1(x_1)E_1(x_2)\psi^\dagger T^a\sigma_i \chi)
 \,a_H^\dag\,a_H\,T(E_1(x_3)E_1(x_4)\chi^\dagger T^a\sigma_i \psi) 
 \mid 0\rangle}
\nonumber
\end{eqnarray}
where
\begin{equation}
a_H^\dag\,a_H  = \sum_X\mid H+X\rangle \langle H+X \mid.
\end{equation}
The leads to the polarization leveling off at large $p_T$ at some
value which is fixed by $R_{M/E}$.  In Fig.~\ref{polar}, we show the
prediction for $J/\psi$ and $\psi'$ polarization at the Tevatron.  The
data is from \cite{Affolder:2000nn}.  The three lines correspond to
different values for $R_{M/E}$=(0 (dashed), 1 (dotted), $\infty$
(solid)). The dashed line is also the prediction for NRQCD$_b$. The
residual transverse polarization for $J/\psi$ at asymptotically large
$p_T$ is due to feed down from $\chi$ states.  The non-perturbative
corrections to our predictions are suppressed by $\lqcd^4/m^4$.

\section{Conclusion}

In this talk I have reviewed NRQCD, and the NRQCD factorization
formalism which is used to make predictions for the production and
decay of charmonium and bottomonium. I did not discuss all of these
predictions. Instead I focused on what I believe to be the most
important prediction of NRQCD factorization: the transverse momentum
distribution of unpolarized and polarized $J/\psi$ and $\psi'$
produced at the Tevatron. Because the unpolarized data can be used to
determine the unknown color-octet matrix elements, it is possible to
make a parameter free prediction for polarized production. This
provides a clean test of the NRQCD factorization formalism. Moreover
the quality of the data for unpolarized production is good, and while
the data for polarized production has large error bars it is expected
to get better.

The NRQCD factorization formalism predicts the $J/\psi$ and $\psi'$ to
be transversely polarized at large $p_T$. This is because at large
transverse momentum, the dominant production mechanism is through
fragmentation from a nearly on shell gluon to the octet $^3S_1$ state.
The quark pair inherits the polarization of the fragmenting gluon, and
is thus transversely polarized~\cite{Cho:1994ih}. The leading order
transition to the final state goes via two $E1$, spin preserving,
gluon emissions. Various corrections dilute the polarization some, but
the prediction still holds that as $p_T$ increases so should the
transverse polarization. Indeed, for the $\psi^\prime$, at large
$p_T\gg m_c$, we expect nearly pure transverse polarization.  The
current experimental results~\cite{Affolder:2000nn} show no or a slight
longitudinal polarization, as $p_T$ increases. If, after the
statistics improve, this trend continues, then it will be the smoking
gun that leads us to conclude that either NRQCD is not the correct
effective field theory for charmonia, or that factorization fails in
these processes.

The possibility that NRQCD is not the correct effective theory for
charmonium leads me to ask: is there any reason to believe that there
is any effective theory to correctly describe the $\jp$?  I believe
that the spin symmetry predictions for the ratio of $\chi$ decays
clearly answers this question in the affirmative.  Assuming that such
an effective theory exists, then is it NRQCD$_c$ or NRQCD$_b$?  As I
have shown the two theories do indeed make quite disparate
predictions, which in principle should be easy to test.  

However, these tests can be clouded by the issues of factorization and
the convergence of the perturbative expansion. One would be justified
to worry about the breakdown of factorization in hadro-production at
small transverse momentum. However, for large transverse momentum one
would expect factorization to hold, with non-factorizable corrections
suppressed by powers of $m_c/p_T$.As far as the perturbative expansion
is concerned, it seems that for most calculations the next-to-leading
order results are indeed smaller than the leading order
result~\cite{Beneke:1997qw,Maltoni:1998nh,NLO}, though, the NNLO
calculation performed, in the leptonic decay width
\cite{Beneke:1997jm}, is not well behaved at this order. 
 
In the end I believe the data will be the final arbiter. The
polarization measurement may fall in line with the NRQCD$_b$
prediction. Or the data may result in longitudinal polarization for
$J/\psi$ and $\psi'$, in which case it may be that neither NRQCD$_c$
nor NRQCD$_b$ are the correct theory.


\begin{theacknowledgments}
I would like that thank my collaborators Adam Leibovich and Ira
Rothstein. Thanks also go the organizers of this conference. This work
was supported in part by the Department of Energy under grant number
DOE-ER-40682-143 and DE-AC02-76CH03000.
\end{theacknowledgments}

\end{document}